# A History and Informal Assessment of the *Slacker Astronomy* Podcast

by **Aaron Price**
American Association of Variable Star Observers
**Pamela Gay**
Harvard University Science Center
**Travis Searle**
American Association of Variable Star Observers
**Gina Brissenden**
University of Arizona

## Abstract

*Slacker Astronomy* is a weekly podcast that covers a recent astronomical news event or discovery. The show has a unique style consisting of irreverent, over-the-top humor combined with a healthy dose of hard science. According to our demographic analysis, the combination of this style and the unique podcasting distribution mechanism allows the show to reach audiences younger and busier than those reached via traditional channels. We report on the successes and challenges of the first year of the show, and provide an informal assessment of its role as a source for astronomical news and concepts for its approximately 15,500 weekly listeners.

## 1. INTRODUCTION

*Slacker Astronomy* is a weekly audio show (podcast) about astronomy. It is distributed on the Internet in MP3 format through the World Wide Web and a new technology called *podcasting*. The goal is to educate audiences, primarily ranging in age from 12 to 50, about a recent news event in the world of astronomy in an entertaining and humorous manner while respecting the intelligence of the audience. In this article, we document our first year's activity and lessons learned. This is one of two papers in a series about podcasting. The other (Gay, Price, & Searle 2006) details how to apply lessons to astronomical podcasting in general.

# 1.1 About Podcasts

Podcasting is a relatively new technology developed in the summer of 2004. It stems from the merger of three technologies: RSS syndication, portable MP3 players, and blogging. When used together, these resources can be used to automatically download and play audio content from the Internet. Referred to as *subscription*, this process is similar to the distribution models of printed newspapers and magazines.

On a schedule set by the listener, which can range from minutes to months, a podcast client on a person's computer checks a list of podcasts to which the listener subscribes. If new content is available, the client downloads the content to the local computer and automatically loads it into his or her MP3 software or onto a connected portable MP3 player.

This process makes the listener's role in content acquisition passive while giving the listener complete control over how and where he or she listens to content. Once the audio file is on a home computer or MP3 player, it can be listened to at leisure (i.e., on demand). An MP3 can be paused, rewound, or fast-forwarded, similar to a television show on a DVR device. This is often referred to as *time shifting*.

Time shifting allows a listener to fit the show into his or her lifestyle. After initially subscribing, listeners do not need to do anything further but maintain Internet access. New shows appear in a listener's audio player as produced. In addition, if a user has a portable MP3 player, the shows can be heard away from the computer. Subscribers do not need a portable MP3 player to listen to a podcast, however; it simply makes it more convenient. Forty percent of *Slacker Astronomy* listeners do not use portable MP3 players. Most people listen to podcasts while involved in daily-life tasks such as commuting to work, exercising, and cleaning. Any Web browser or computer produced in the last three years should include built-in support for playing MP3s.

Podcasting has experienced phenomenal growth because of this ease of use. The two largest podcast directories are Podcast Alley (podcastalley.com), with over 17,400 podcasts, and Apple's iTunes service, which boasts 25,000 podcasts in its directory (Apple 2006). The popular press has covered podcasting extensively as an alternative (and threat) to radio— alongside satellite radio, which simulcasts many podcasts, including occasional episodes of *Slacker Astronomy* (Sirius Satellite Radio 2005).

Finally, an added advantage of podcasts is that the show is available as a permanent record to the listener. If a listener likes a show or would like to share it with others, he or she can easily keep or transfer the MP3 file that is on his or her hard drive. This makes the show's content available to listeners as a reference as well.

The Pew Internet and American Life Project released a survey of 2,201 Americans in April 2005 (Rainie & Madden 2005). They estimate that over six million Americans had listened to at least one podcast, and 29 +/- 7.8% of the owners of portable MP3 players had listened to podcasts. And this was before Apple added podcast support to iTunes.

According to Jupiter Research (2005), 22 million Americans owned portable MP3 players in 2005, and sales continue to climb. MP3 players are increasingly common among our target demographic. Twenty percent of those between 18 and 28 years of age own MP3 players. Jupiter Research predicts that 56.1 million portable MP3 players will be sold in the United States by 2008. If 30% of users listen to podcasts—a conservative number based on the current rate—then the potential market for *Slacker Astronomy* is nearly 17 million listeners.

## 2. ABOUT *SLACKER ASTRONOMY*

## 2.1 History

In December 2004, podcasting began to receive attention in the mainstream media. The first author (A. Price) came across a newspaper article about the proliferation of religious programming in this new medium. Knowing that podcasting served an early adopting, tech-savvy audience, he believed that an astronomy show would be a hit. The hook was to present the content to the audience on their own terms, which meant combining serious science with a presentation that was anything but serious. Price brought in P. Gay and T. Searle to help, and the show has been a collaboration of all three from the very beginning.

On February 14, 2005, the first show went online. As of April 4, 2006, 48 scripted episodes have been produced, along with 33 unscripted bonus episodes. *Slacker Astronomy* has about 15,500 unique subscribers who download at least one show per week, and about 2,700 weekly subscribers who listen to the unscripted episodes on a bonus feed that we call SA-Extra. The slackerastronomy.org Web site receives 19,500 unique visitors per week. About 20 e-mails are received per week, most with questions about an astronomical topic that we covered in a show, and about 20 posts per week are made in discussion forums on the Web site.

*Slacker Astronomy* was the first science podcast, according to listings at ipodder.org, the predominant podcast directory prior to the launch of the iTunes Music Store directory. In fact, ipodder.org created its science category for *Slacker Astronomy*, which was originally listed under Current Affairs (!). Science@NASA was also online at the time but did not list itself in podcasting directories (we speculate that this is because the show was a reading of the Science@NASA Web page and did not contain original content, which was considered part of the definition of a podcast at the time).

*Slacker Astronomy* owes a part of its success to another podcast called *The Daily Source Code*, hosted by former MTV VJ Adam Curry. Curry coinvented the podcast technology and concept, and his show is one of the most popular podcasts, with over 120,000 subscribers. Unsolicited, he listened to our fourth show, which aired on February 20, 2005. In the February 24, 2005, episode of his daily podcast, he included a lengthy and positive review of *Slacker Astronomy* (Curry 2005). Within hours, our servers became overloaded as listeners subscribed to our show en masse, driving our listenership from 140 to 2,000 within a week.

This gave us the critical mass of listeners needed to make a splash on the scene. A week later, on March 2, 2005, MSNBC.com featured another positive review of our show on the home page of their Space section. This resulted in 500 to 1,000 additional subscribers (Boyle 2005). A month after that, WGBH-FM, a public radio station in Boston, interviewed us for an edition of their own podcast, *Morning Stories* (Kahn 2006), which plays over the air on their traditional radio station. Since then, we have been included in on-air broadcasts on National Public Radio (NPR) in May 2005 (Montagne 2006) and the British Broadcasting Corporation (BBC) in August 2005 (BBC Radio Channel Five 2005). In addition, shows are occasionally aired on Sirius Satellite Radio (Curry 2006).

The podcasting community exploded in popularity and size with the release of Apple's iTunes 4.9 in July 2005. This version of Apple's popular music player and store added podcast subscription features, including a directory. Within a month, *Slacker Astronomy* landed in the iTunes Top 100, peaking at 59. In addition, the iTunes staff featured *Slacker Astronomy* on its podcast home page as a "recommended indie podcast" from August 4 to August 11, 2005. From October 13 to October 19, 2005, the iTunes podcast store featured *Slacker Astronomy* on its home page, this time under "Recommended Science Podcasts." During this period, our show again reached their top 100, peaking at 28 (Figure 1). In January 2006, the American and British versions of the iTunes Music Store recommended *Slacker Astronomy* in the Comedy category. These were all unsolicited promotions of our show by the iTunes staff, with whom we have had no contact. In our experience, the iTunes Store usually features commercial podcasts and Apple business partners, which makes it very difficult for independent podcasts like ours to get front-page promotion. This is an ongoing complaint and discussion in popular podcaster forums such as PodcastAlley.com (Podcast Alley 2006). Apple has over 25,000 podcasts in its directory from which to choose and has chosen to feature us many times.

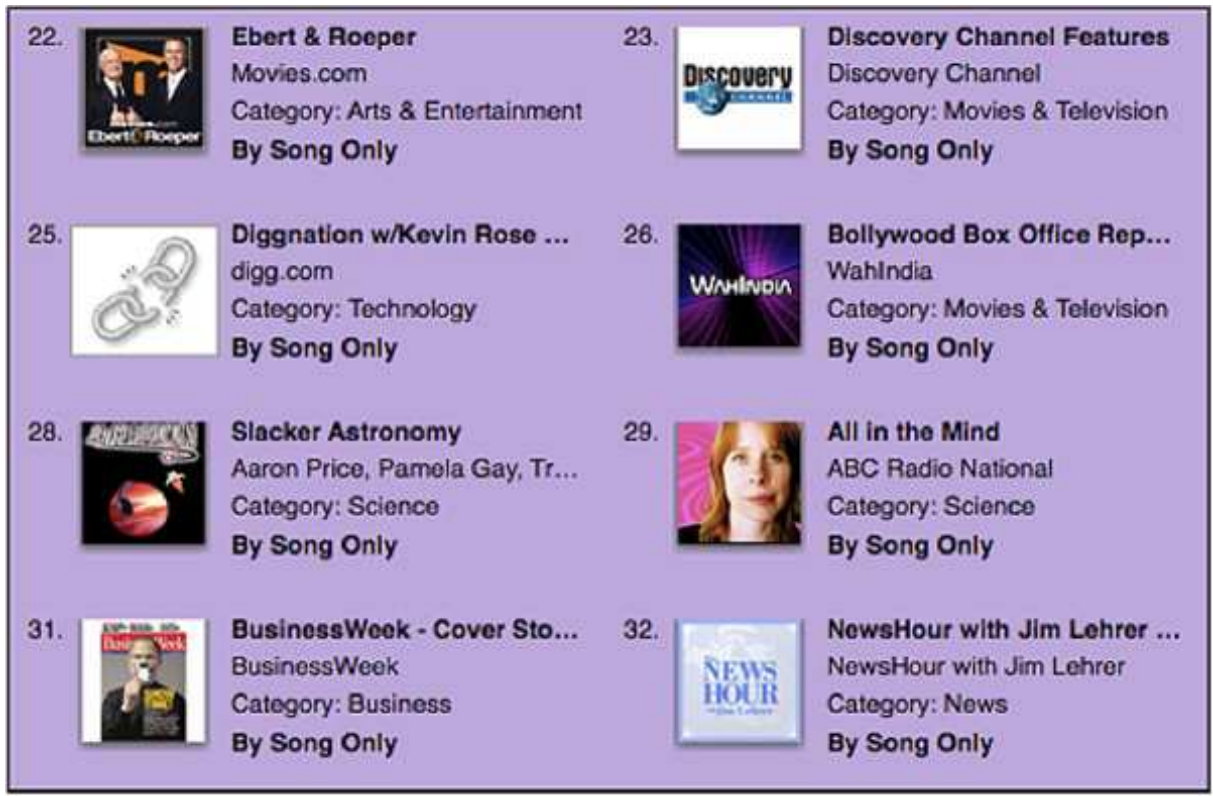

**Figure 1.** A Clip of the iTunes Top 100 Podcast Listing on October 17, 2005. (*Slacker Astronomy* is 28 out of 25,000. Notice the type of the other podcasts, all commercial.)

In early 2006, Apple's iTunes began keeping track of the top shows per category. Since then, *Slacker Astronomy* has been consistently in the top 10 listing of the Science category, peaking at #4 (Figure 2). Of the science podcasts, *Slacker Astronomy* is first among those with original content. The higher-rated podcasts are redistributions of science content first published in other places (Web sites, magazines, radio, and so on). In addition, iTunes allows listeners to post reviews and ratings of podcasts. Out of 34 posted reviews, *Slacker Astronomy* has an average rating of four out of five stars.

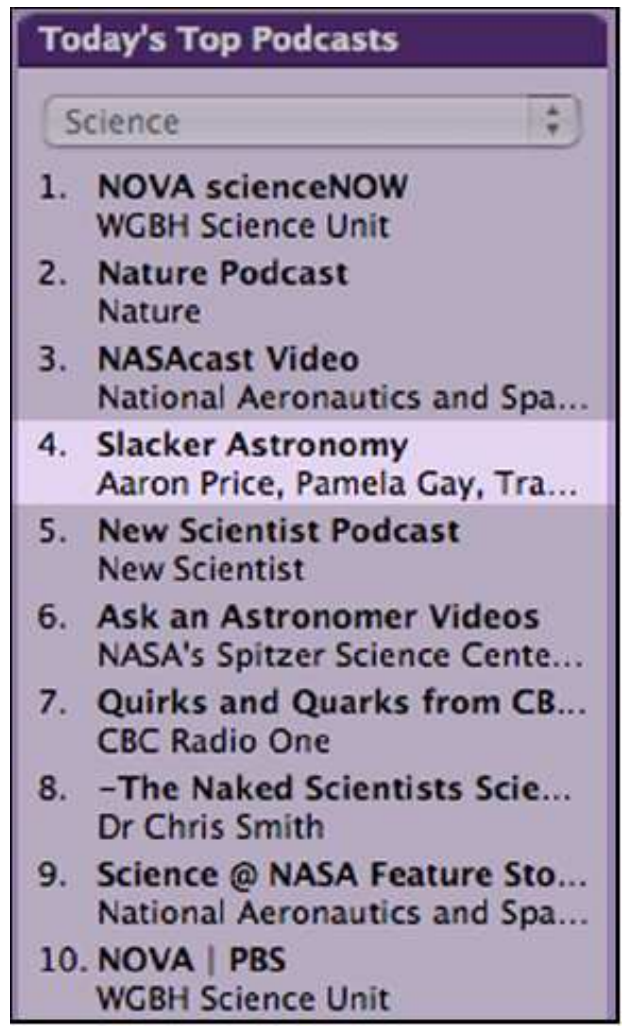

**Figure 2.** A Clip of the iTunes Top 100 Podcast Listing for the Science Category on January 4, 2006

## 2.2 Current Production Status

A. Price is the project manager and head writer, contributing two thirds of the scripts, and he puts together the majority of the Bonus Features. P. Gay writes the balance of the scripts, edits all scripts, and does the balance of the Bonus Features. T. Searle and P. Gay share the on-air hosting duties. Searle is also responsible for audio engineering, including special effects and music composition, and he directs all Chit Chat shows. Everyone is capable of handling anyone else's duties, which supports a flexible production schedule. The show is currently produced entirely on volunteer time by the three founding members, with no official support from their employers. On average, about 15 combined person-hours per week are invested into producing and supporting the show.

Currently, hosting and bandwidth costs run about $150 a month, an amount that is rising in direct proportion to audience growth. The fees are currently covered by listener donations, which we project will continue to the end of 2006. We secured commercial sponsorships to fund trips to astronomical and

teaching conferences by members of the team and to prepare for the day when listener donations can no longer cover the bandwidth fees. Most equipment and software have been provided by the team members themselves. The team is looking for other means that will offer long-term support and allow the team to spend more time on the project.

We also reached an agreement with the Mobile Broadcast Network, a contractor of Nextel Communications, to offer *Slacker Astronomy* episodes within the next six months to those who own mobile phones on the Nextel network. Negotiations are also under way with other possible distribution channels.

Episodes of *Slacker Astronomy* have been downloaded 1,044,325 times as of April 3, 2006. The average listener downloads around 50 shows, which suggests that listeners like the first show they hear enough to download the rest. (Most podcast players, such as iTunes, do not automatically download all shows; they usually only download the most recent.) In addition, unlike a TV or radio show, these episodes may be permanently available on the Internet. Thus, downloads should continue into the foreseeable future. Our hope is that *Slacker Astronomy* will evolve alongside the latest technological advances of the time. At the very least, these podcasts should always be available to anyone with Internet access. It is likely that the content will be outdated before the technology is.

## 2.3 Content

Being early to market and without precedent, *Slacker Astronomy* was free from the limitations of convention, and we were allowed to explore our own path to efficiently produce podcasts, using them to communicate science to audiences who normally do not follow it. Our original content goals were to (1) entertain, (2) provide astronomy news and information to those who do not otherwise follow it, (3) introduce stories not covered in mainstream science outlets, and (4) introduce and explain advanced concepts.

*Slacker* is a relatively new term in the English lexicon (Merriam-Webster 2006). It refers to a person, generally young, who not only underachieves but is also proud of it. It became the self-imposed label of generations X and Y after the success of the 1990 independent movie of the same name. We instilled this attitude in our online personas in a humorous and ironic way while communicating astronomy. By embracing this attitude, we bring a sense of levity to a topic too often presented as serious and grandiose. Instantly, listeners know what they are going to get from the title alone. Feedback from listener e-mails reveals that many listeners are first attracted to our show over others largely because of its title.

Our shows are a blend of simple comedy and news reporting. The two hosts, Searle and Gay, banter back and forth about a recent news event in astronomy. They joke, include pop culture references, and sometimes role-play humorous short sketches about the topic. They have even been known to sing on occasion. On average, we estimate that 10%–15% of the show is comedy, but even in the rest of the show, we address the audience as if we were longtime friends. Our aim is to bypass stereotypes of astronomers. In Curry's (2005) review of the show, he said, "Don't you just wish you had a professor who talked about astronomy in this way? . . . it's just beautiful." The tone is casual, the pace is conversational, and the vocabulary is full of colloquialisms, all carefully crafted. Unlike the vast majority of podcasts, which are informal, often unstructured, conversations, our shows are scripted, and the scripts are available on our Web site when the show is released.

Our levity would not work if not for the other aspect of our unique style: respect for the audience's intelligence and curiosity. We are not afraid to include hard science or background lessons so that the audience understands the significance of the news event being covered. For example, in one of two shows about aurorae (February 28, 2005), we specifically explained the ionization process by which photons are emitted by the oxygen and nitrogen atoms in the upper atmosphere. Popular science often leaves out these details, yet the online community thirsts for them. We have come to believe, based on listener feedback, that modern, Internet-savvy listeners have a ravenous appetite for details. They do not want an easy-to-understand explanation of a phenomenon or a sound bite. They express a genuine curiosity, wanting to know why and how. The original plan was for the show to remain under five minutes, within the attention spans of an audience accustomed to listening to musical songs on their MP3 players. However, the most common early complaint of our show was that it was too short (and too infrequent), so we have increased the average length of our shows to about nine minutes. This kind of response to audience feedback goes a long way in building listener loyalty.

Topics for the shows come from many sources, beginning with our daily perusal of astronomical journals, the ArXiv scientific paper preprint server, and press releases issued through the American Astronomical Society (AAS) press office (Price and Gay both have AAS press credentials). The latter is an e-mail list clearinghouse of almost all astronomy-related press releases. We focus each show on a recent news event or series of related events, using discoveries to teach both fundamental and advanced astronomical concepts. Using current events keeps the show topical and fresh, as opposed to rehashing the "top 20" of astronomical subjects, which the listeners may have heard many times over their lifetimes. Listeners are still introduced to those topics, but in the context of a new discovery. Because all three team members work in the field of astronomy, we are able to catch and cover topics that are important but usually ignored by the mainstream science press. For example, instead of covering Deep Impact at the time of collision, we waited two months and produced an episode explaining the science that later emerged in professional journals.

In addition to the regular weekly shows, we publish special bonus episodes on a subscription feed that we call SA-Extra. The SA-Extra episodes usually are longer (30–60 minutes) and are unscripted. Here is a sampling of SA topics: interviews with other professional astronomers, "sound seeing tours" of astronomical conferences, facilities, and star parties, public lecture recordings, and Q&A shows during which we answer listener e-mail and give updates on past news stories. Our goal with this feed is to give more depth while also providing a behind-the-scenes look into astronomy. Listeners especially enjoy our profile interviews with professional astronomers.

A pleasant surprise has been the number of teachers and parents who play the show for children. We learned of this through e-mail feedback. In response to their e-mail requests, we created a rating system that identified family-friendly shows with an "FF" in the title. We also use "OF" for "office friendly" shows and "JF" to describe shows made "just for fun." Since the rating system began, 75% of our shows were rated FF, 20% OF, and 5% JF.

Our Web site is a critical part of the program. Not only does it serve as the distribution point, but it also allows our listeners to learn more on a subject and build a community. For every episode, we post Show Notes, which include links to original research papers or press releases, links to background information, images, and diagrams, a full transcript of the show, and more. It also is a place for listeners to ask follow-up questions about the material. For the 207th meeting of the American Astronomical Society, we included a very popular blog that we updated many times per day with news and reports from the meeting floor. Over time, a community has grown around the show among the 80,000 or so unique monthly visitors.

Astronomy is a gateway to critical thinking and imagination. Through these podcasts, we attempt to expose the listener to concepts such as skepticism and wonder. We prefer to do it by explicitly exposing both the true meaning of a discovery and the scientific process that went into revealing it. These concepts alone can sometimes seem vague, but within the context of astronomy, they become approachable. Younger minds are very impressionable, which is why marketing agents crave the 18–50-year-old audiences (Koschat & Putsis 2000). By keeping astronomy on the radar for the generations in this age group, we keep it from being forgotten or marginalized. We believe that it is much easier to keep someone interested than it is to rekindle a dormant interest later in life. This goes beyond the podcast-to-listener connection and multiplies in the online community. Boundaries are blurred on the Internet as dialogues begun on our Web site spread to listener blogs, Web sites, and other podcasts. Often we find our shows discussed and linked on listeners' sites and in their message boards. We especially enjoy seeing discussions of our shows in foreign languages. This creates a value-added awareness of astronomy as concepts in a particular show expand further across the World Wide Web.

## 3. INFORMAL IMPACT ASSESSMENT

## 3.1 Demographics

Because podcasting is new, it has been a challenge to quantify our results. E-mail and Web site discussions have been our main channel for judging the response to the show. In September 2005, at the 117th meeting of the Astronomical Society of the Pacific, P. Gay gave a talk, "Reaching the Podosphere," at a conference titled "E/PO as an Emerging Profession." A demographic survey was conducted in August 2005 to support the talk. About 428 surveys were downloaded and 347 responses submitted. Details are available in Gay, Price, and Searle (2006) and mentioned here in support of an evaluation conclusion.

In the August 2005 survey, we asked the listeners to estimate their level of involvement in astronomy before and after they began listening to the podcast (Table 1). We gave them two choices, one describing a passive interest in astronomical news and the other describing an active interest. The results reflect an increase of initiative among listeners in seeking out astronomical information.

**Table 1.** Choice answers and responses to the survey questions, "How would you describe your interest in astronomy [prior to/after] listening to astronomy related podcasts?" We received 232 responses to this question out of 592 +/- 10% survey downloads.

| **Before** | **After** | **Self-Estimated Level of Involvement** |
|---|---|---|
| 44% | 18% | I will pay attention if it crops up in something I already read/watch/listen to. |
| 56% | 82% | I actively seek and read/watch/listen to stories about astronomy. |

In our observation, the listeners of podcasts can be described in one word: busy. Podcasting appeals to busy people in the same manner as VCRs and TiVos. The listener is in complete control of where, when, how, and to what he or she listens. As a result, *Slacker Astronomy* has an opportunity to reach a unique audience unable to get astronomy news through traditional channels because of their broadcasting constraints. According to the August 2005 survey, approximately 34% of our surveyed listeners do not actively seek out astronomy-related news or information outside our podcast. Through our correspondence with the audience, we find that many listeners are teenagers and young adults who are often too busy to obtain astronomy content in any way other than our show, yet they have an interest in astronomy. They simply need a way to get the information that fits their lifestyle. According to an attitudinal survey discussed next, about 85% of the surveyed *Slacker Astronomy* listeners perform other tasks while listening to the show.

## 3.2 Informal Assessment

Beginning in December 2005, we conducted a series of interviews, attitudinal surveys, and knowledge surveys to assess our effectiveness in education and public outreach.

The first stage involved interviews with five subjects to determine outstanding needs of our listeners. We divided our listeners into five categories following a distillation process developed from the quality function deployment (QFD) process. QFD is a well-respected methodology that has been in use for decades and is best described in the book of the same name (Cohen 1995). In the distillation process, we brainstormed every possible category of listener we could concieve. Then we categorized them into the smallest number of groups that included each type of listener. The goal of this distillation was to come up with the smallest of number of groups that would encompass the needs of every listener. A member of each group was interviewed for 30–45 minutes via telephone or Internet telephony. They were asked about their astronomical interests, history, and desires. The interviews were recorded and analyzed later to determine the needs.

Although QFD is an effective methodology for product development, we determined that it is not an effective methodology for our product evaluation. Most of the interview subjects were satisfied with the way they receive astronomical news. This is certainly a selection effect, because we interviewed active listeners and not those who had unsubscribed or who had chosen not to subscribe in the first place. Although we did learn more about our audience and came away with some interesting ideas for show topics, we did not get to the core needs of our audience. A more effective interview technique must be determined for our next needs-analysis evaluation, and we need to put more effort into choosing subjects outside our listenership.

We found more success with attitudinal and knowledge surveys developed with the Field-Tested Learning Assessment Guide (FLAG) primer for attitudinal assessment (Zeilik & Mathieu 2000; Brissenden, Slater, Mathieu, & NISE 2002). A 19-question survey was placed on our Web site from December 27, 2005, to January 24, 2006 (Table 2). Questions were chosen to both help determine listener behavior and to address each category of Bloom's Taxonomy for Affective Goals (Bloom & Kratwohld 1956; Kratwohld, Bloom, & Masia 1964). The taxonomy is designed to characterize, from simple to complex, the different levels of a learning process involving growth in feelings or other emotional areas. Responses are placed into a forced-choice Likert scale (Likert 1932) using an even number of categories to offset the tendency of survey takers to choose the middle. We received 465 responses from December 27, 2005 to January 25 , 2006. Responses were on an ascending scale of 1 to 6, from *strongly agree* to *strongly disagree*, except when the scale ascended from *always* to *never*, indicated by an asterisk. Scores to † were reversed for placement on the same ascending scale.

**Table 2.** Questions and average scores from the attitudinal survey conducted from December 27, 2005, to January 25 , 2006. A total of 465 responses were received out of an estimated 592 (+/- 10%) downloaded surveys. Responses were on a Likert scale of 1–6, ascending from *strongly agree* to *strongly disagree*, except when the scale ascended from *always* to *never*, indicated by an asterisk. Scores to † were reversed for placement on the same ascending scale.

| Question | Category of Bloom's Taxonomy | Average Score (SD) |
| --- | --- | --- |
| 1. The scripts are easy to understand. | 1 | 1.78 (0.10) |
| 2. I feel like I understand the concepts in the shows. | 1 | 1.62 (0.12) |
| 3. I stop listening to a *Slacker Astronomy* show before it is over.* | 1 | 1.04 (0.18)† |
| 4. I skip through portions of a Slacker Astronomy show before it is over.* | 1 | 1.81 (0.16)† |
| 5. The show notes on the Web site are helpful to me. | 2 | 2.73 (0.12) |
| 6. I have spoken to other people (friends, family, coworkers, etc.) about things I have heard on the show. | 3 | 2.88 (0.12) |
| 7. A fact or concept I heard on the show surprised me.* | 4 | 2.16 (0.14) |
| 8. I have sought further information about a topic I heard on *Slacker Astronomy*. | 4 | 2.90 (0.14) |
| 9. I invest more time in astronomy activities than I did before listening to *Slacker Astronomy*. | 5 | 3.13 (0.14) |
| 10. I listen to a show more than once.* | N/A | 3.41 (0.14) |
| 11. I like the choice of topics in the shows. | N/A | 2.64 (0.13) |
| 12. I doubt the validity of the information I hear on the show. | N/A | 5.37 (0.16) |
| 13. When a show is over, I am often confused about what I just heard. | N/A | 4.70 (0.16) |
| 14. I consider *Slacker Astronomy* educational. | N/A | 1.74 (0.10) |
| 15. I consider *Slacker Astronomy* humorous. | N/A | 2.02 (0.09) |
| 16. I prefer the more humorous shows. | N/A | 2.81 (0.14) |
| 17. I prefer the more serious shows. | N/A | 3.32 (0.20) |

By categorizing our survey questions into Bloom's Taxonomy, we were able to come up with an index for the effectiveness of our show, with 1 being most effective and 5 being least effective (see Table 1 and Figure 3). We took the average if multiple questions were assigned to the same category (we reversed the score for questions 15 and 16, which had a reversed scale in how the question was phrased).

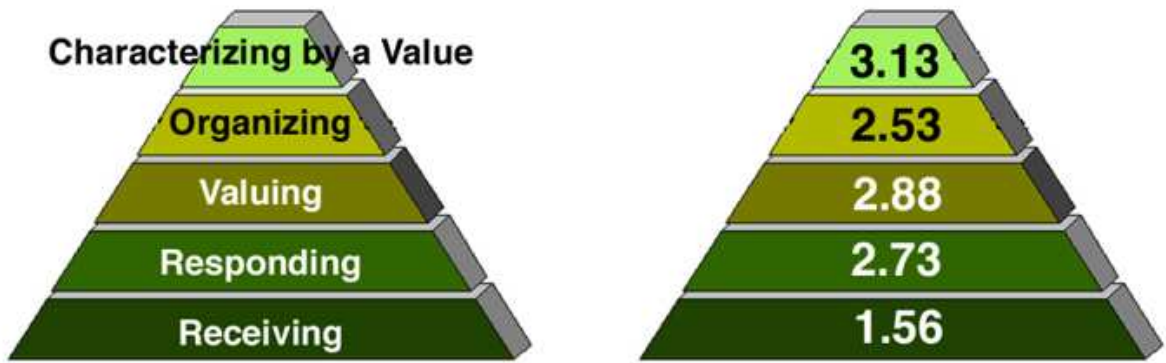

**Figure 3.** Average Categorized Attitudinal Survey Scores (right) Placed into the Context of Bloom's Taxonomy of Affective Goals (left)

In addition to the interviews and the attitudinal surveys, we conducted two contextual surveys to test the retainment of facts (see Table 4 a; Table 5). The first knowledge survey—197 responses out of an estimated 592 (+/- 10%) downloaded surveys—was about a show we aired approximately one month earlier. Although we cannot assume when subscribers listened to a show, feedback suggests that the majority typically listen within a week or two. The show addressed the measurement of Saturn's rings by Cassini scientists using spectroscopic observations of the star Mira as it passed behind the rings. The second knowledge survey—126 responses out of an estimated 478 (+/- 10%) downloaded surveys—covered a show we aired approximately two weeks earlier. Its content was based on a summary of discoveries made by the Huygens probe as it landed on Titan.

It is difficult to judge the response rate for online surveys. We had three main methods of marketing our surveys. First, mentioned it in our show. Second, we linked to the survey from our Web site. Third, we mentioned it on one of our bonus shows on the SA-Extra feed. Because of our production schedule—shows are sometimes recorded far in advance of the publishing date—we were unable to consistently use all three methods. For example, the attitudinal survey was mentioned via all three methods, but the knowledge surveys were not. Therefore, we defined the total number of surveys downloaded as the number of times the survey Web page was downloaded by a unique IP address. In cases of multiple surveys per IP address, we processed only the last one submitted. Therefore, multiple downloads by one IP address counted as only one survey download in our report. We believe that the total number of downloads of each survey to be conservatively accurate to be within 10%, based on the percentage of IP addresses that downloaded the survey more than once.

Each survey included four questions about a particular show. Three questions were about a science fact presented on the show and included multiple-choice answers. The fourth question asked responders to explain a fact or concept introduced in the show. Responses were categorized based on whether they had listened to the show (control). In both surveys, we found that the number of correct answers from among those who listened to the show was higher than the number of correct answers from those who did not listen to the show. The second survey asked more technical questions than the first. As expected, the difference between the listeners and the control group increased (Figure 5). We will rerun these exact survey questions in approximately six months to gauge how much of the material has been retained.

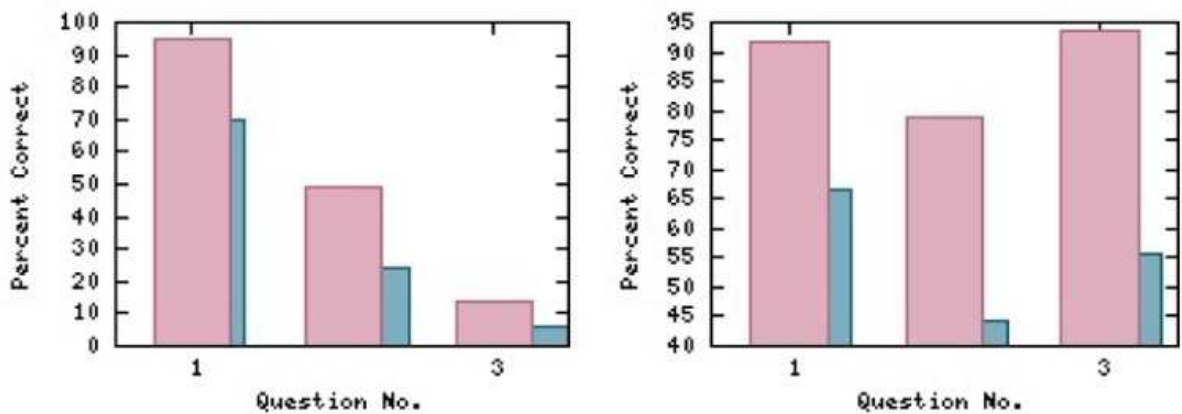

**Figure 4.** Percentage of Correct Answers to Two Knowledge Surveys. (Red represents listeners to the show, and blue represents nonlisteners.)

**Table 3.** The first knowledge survey, 197 responses out of an estimated 592 (+/- 10%) downloaded surveys, was about a show that we aired approximately one month earlier. Although one cannot assume that subscribers listened to it then, feedback suggests that the majority typically listen within a week or two. Answers are asterisked.

| Question | Answer Choices |
|---|---|
| **When a star passes behind an object, it is called a(n):** | Occultation* |
| | Eclipse |
| | Observation |
| | None of the above |
| **Saturn's rings are on the order of ____ thick when seen edge-on.** | Meters |
| | Tens of Meters* |
| | Hundreds of Meters |
| | Thousands of Meters |
| | Don't Know |
| **Over how long of a period did Cassini image the star Mira?** | Minutes |
| | Hours |
| | Days |
| | Weeks |
| | Months* |
| | Don't Know |

**Table 4.** The second knowledge survey, 126 responses out of an estimated 478 (+/- 10%) downloaded surveys, covered a show that we aired approximately two weeks before publication of the survey.

| Question | Answer Choices |
|---|---|
| **The surfaces of Earth and Titan both have very similar:** | Meteorology, geomorphology, and fluvial activity* |
| | Viscosity, gravity, atmospheres |
| | Chemistry, carbon-14 ratios, and atmospheres |
| **A methanogen is:** | A type of rock |
| | A type of bacterium* |
| | A type of river |
| **While descending to the surface of Titan, the Huygen's probe imaged:** | Active volcanoes |
| | Granite outcroppings |
| | Methane rivers* |

# 4. CONCLUSIONS AND THE FUTURE

## 4.1 Video Expansion

Expansion into video production will take place in 2006. In the summer of 2005, the Don Harrington Discovery Center in Amarillo, Texas, approached *Slacker Astronomy* about licensing our scripts to produce shows for their new planetarium theater. Although all *Slacker Astronomy* material is available under the Creative Commons Attribution-NonCommercial-ShareAlike 2.5 License (Creative Commons Deed 2006), we decided to work together to develop a show based on new content so that we may customize and affect how our style will translate visually. The show will take the *Slacker Astronomy* style and repackage it into a visual presentation using the state-of-the-art Digistar 3 digital planetarium system, which has been installed in 43 planetariums worldwide. Its predecessor, the Digistar II system, has been installed in 120 planetariums. The need for content will grow as this community upgrades to the Digistar 3 system. The only content currently available for it consists of packages generated by the manufacturer. These packages tend to focus on pretty pictures and are very light on science, based on our experience with them and audience feedback reported by the Don Harrington Discovery Center. For these presentations, the vendor charges an initial distribution fee, not publicly disclosed, and also requires 25% of the ticket revenue ("gate"), which must be reported to them in quarterly installments. This is significant cost and overhead for most planetariums, especially smaller and rural ones. Our podcast will provide alternative, lower-cost content.

In addition, the release of the full-color video iPod has allowed many podcasts to add video elements to their shows. Thus far, available material consists mainly of redistributed video originally developed for television. We plan to create unique customized video content meant only for the podcast audience. The show may use many of the technical elements from the planetarium show, allowing us to save production time by developing both simultaneously. A video podcast about a photon's journey out of the Sun is in preliminary development and planned for release in mid-2006.

These new endeavors will not be without their challenges. Video production has significant technical and financial challenges, and video serves a different audience than podcasting does. This first video podcast and planetarium show are but experiments to determine the feasibility of expanding the *Slacker Astronomy* brand.

One of the recent survey questions we asked was, "Are you interested in other science shows with the same format and style?" About 98% of the respondents chose Yes. This illustrates that *Slacker Astronomy*'s formula for success need not be restricted to podcasts and astronomy. In the next year, research will be conducted into possible expansion of the podcast into other categories such as *Slacker Mathematics* and *Slacker Physics*.

## 4.2 Conclusion

The *Slacker Astronomy* show attempts to unearth the wonder of astronomy from beneath layers of stereotypes and preconceptions. We entertain around 15,500 listeners each week with astronomy-related schtick—heavy on the cheese, heavy on the science. Surveys show that almost all listeners come away with new ideas and knowledge of astronomy, and most share this knowledge with others. This was the easy part. The next challenge will be to obtain better penetration of the two deepest levels of Bloom's Taxonomy: organization and characterization by a value.

We must also establish effective outreach goals. The team originally started this collaboration for personal fun. Initially, everything was done with only that in mind. Along with the unexpected success comes an opportunity to more carefully consider the education and public outreach goals of *Slacker Astronomy*. To more efficiently take advantage of the opportunity, we used the results of these assessments to develop a mission statement to guide our second year of programming (Figure 5).

> The Slacker Astronomy podcast and web site introduces recent discoveries and news events from the world of astronomy. The goals of the podcast are:
> 1. To have fun and show it
> 2. To demonstrate that astronomy is part of everyone's daily universe
> 3. To judge each discovery or news event by its own merits
> 4. To aid our audience in overcoming scientific preconceptions and misconceptions

**Figure 5.** The New *Slacker Astronomy* Mission Statement

The popularity of the *Slacker Astronomy* podcast has far exceeded the expectations of its producers. Although we focused mainly on entertainment during the first year, the second year will include a more formalized approach to astronomical education and public outreach.

## Acknowledgments

We would like to thank our listeners for supporting the bandwidth costs of the show and providing us with copious amounts of feedback. We would also like to thank Phil Plait for early advice to help get us started and Arne Henden for providing moral support.

ÆR
Innovation